\documentclass[aps,prd,prl,reprint,groupedaddress,floatfix,longbibliography]{revtex4-2}

\usepackage{graphicx}
\usepackage{xcolor}
\usepackage{bm}% bold math
\usepackage{physics}
\usepackage{amsmath, amssymb, amscd, amsthm, amsfonts}
\usepackage{braket}
\usepackage{soul}

\begin{document}

%\preprint{}

%\title{How to learn efficient swimming in a nonuniform flow field.}
\title{Learning to swim efficiently in a nonuniform flow field}


\author{Krongtum Sankaewtong$^1$}
\email{aom@cheme.kyoto-u.ac.jp}
\author{John J. Molina$^1$}
\author{Matthew S. Turner$^{2,1}$}
\author{Ryoichi Yamamoto$^{1}$}
\email{ryoichi@cheme.kyoto-u.ac.jp}
%\thanks{}
\affiliation{
$^1$Department of Chemical Engineering, Kyoto University, Kyoto 615-8510, Japan\\
$^2$Department of Physics, University of Warwick, Coventry CV4 7AL, UK
}


\date{\today}
\hfill
\begin{abstract}
Microswimmers can acquire information on the surrounding fluid by sensing mechanical queues. They can then navigate in response to these signals. We analyse this navigation by combining deep reinforcement learning with direct numerical simulations to resolve the hydrodynamics. We study how local and non-local information can be used to train a swimmer to achieve particular swimming tasks in a non-uniform flow field, in particular a zig-zag shear flow. The swimming tasks are (1) learning how to swim in the vorticity direction, (2) the 
shear-gradient direction, and (3) the shear flow direction.   
%MST global->lab frame throughout --> (A) Fixed
We find that access to lab frame information on the swimmer's instantaneous orientation is all that is required in order to reach the optimal policy for (1,2). However, information on both the translational and rotational velocities seem to be required to achieve (3). Inspired by biological microorganisms we also consider the case where the swimmers sense local information, i.e. surface hydrodynamic forces, together with a signal direction. This might correspond to gravity or, for micro-organisms with light sensors, a light source. In this case, we show that the swimmer can reach a comparable level of performance as a swimmer with access to lab frame variables. We also analyse the role of different swimming modes, i.e. pusher, puller, and neutral swimmers. 
%We establish that deep reinforcement learning along with the direct numerical simulation to study swimming in complex fluids.
\end{abstract}

%\keywords{}

\maketitle

%\section{}
%\subsection{}
%\subsubsection{}

% If in two-column mode, this environment will change to single-column
% format so that long equations can be displayed. Use
%\begin{widetext}
%\end{widetext}

% figures should be put into the text as floats.
% Use the graphics or graphicx packages (distributed with LaTeX2e)
% and the \includegraphics macro defined in those packages.
% See the LaTeX Graphics Companion by Michel Goosens, Sebastian Rahtz,
% entire page. There is no need to do explicit centering.

% \begin{figure}
% \end{figure}

% Surround figure environment with turnpage environment for landscape
% figure
% \begin{figure}
% \end{figure}
\section{Introduction}
Active matter encompasses a broad range of physical, chemical and biological systems composed of ``active'' agents that consume energy from the surrounding environment in order to perform tasks (e.g., self-propel). Examples include motile cells such as spermatozoa, fish, birds, and even humans. Besides consuming energy, living agents also sense and react to environmental stimuli in order to accomplish their tasks, e.g., biological objectives such as gravitaxis \cite{Hader2017,tenHagen2014}, chemotaxis \cite{BERG1972,deOliveira2016}, or predation avoidance \cite{Gemmell2014,Michalec2015}. An example of predator avoidance is 
given by copepods (\textit{Acartia tonsa}) \cite{Kiorboe1999}, crustaceans that can be found in both freshwater and saltwater, which use mechanoreceptors to sense hydrodynamic signals in order to escape from predators. The recent progress in synthetic active particles has also revealed exciting possibilities for novel applications of active systems, e.g. as micromotors \cite{C3TB21451F,Karshalev2018} or for therapeutics \cite{FERNANDES20121579,deavila2017}. However, such applications require that the active agents be able to navigate complex environments in order to accomplish particular tasks. A natural question here is how to efficiently train the agents to achieve these objectives, when they are only able to process simple cues from their surroundings. For wet active systems, the major challenge is to fully account for the hydrodynamic interactions. Colabrese \textit{et al.} \cite{Colabrese2017} have used a reinforcement learning method (Q-learning) to develop efficient swimming strategies for a gyrotactic microswimmer, which was tasked with swimming in the vertical direction against a periodic Taylor-Green background vortex flow in two-dimensions. The swimmer was given information of its lab frame orientation and the vorticity of the background flow, which was discretized to have three possible values: positive, negative, or zero. Subsequent studies have extended the method to three-dimensions, e.g., to optimize for vertical migration against gravity \cite{Gustavsson2017,Qiu2022}. However, the hydrodynamic interactions were not fully taken into account in these studies, as the background flow was fixed, with the swimmer being advected/rotated by the flow. In contrast, here we investigate how to train a swimmer to navigate a complex flow (a zig-zag shear flow) by performing Direct Numerical Simulations (DNS) \cite{Yamamoto2021} to account for the full hydrodynamic interactions and the particle-fluid coupling in three-dimensions. First, we consider the case where the swimmer is able to perceive its current location, orientation, and translational and rotational velocities, within the lab frame, as well as retaining a memory of the last two actions it has performed. We then train the swimmer to achieve three separate tasks, swimming in (1) the vorticity direction, (2) the shear gradient direction and (3) the flow direction.  We employ Deep Q-learning \cite{vanhasselt2015} on a suitably discretised action space. Our results show the feasibility of using only the orientation and the action memory in order to learn optimal swimming strategies for tasks (1) and (2), i.e., swimming along the vorticity and the shear gradient directions. Swimming in the flow direction 
%MST reserve this to later discussion, if at all:
%, where the ``do nothing and allow yourself to be advected by the flow'' is the optimal decision, 
proved to be a much more challenging task, as evidenced by the low performance compared to that of the other two. In this case, the swimmer was unable to learn to align itself with the flow streamline. 
%, as it had no memory of where it were at previous times.  %mention this later
While most studies on navigation \cite{Zhu2021,Colabrese2018,muinoslandin2018} assume the agent's state to be comprised of lab frame information, this is not appropriate for biological micro-swimmers, as they can only sense local information, e.g., hydrodynamic signals. Therefore, we have also investigated how the same learning can be performed using only locally accessible information, i.e., the hydrodynamic force exerted on the swimmer by the surrounding fluid and the relative alignment of the swimmer with a signal direction. In what follows we will refer to this as a light source, sensed by light sensitive receptors for which we use the shorthand of ``eye'', without implying the presence of a fully developed eye. In principle micro-organisms could also sense the earth's gravitational field, or other signals coming from e.g. persistent magnetic \cite{MONTEIL2020266}, heat\cite{Almblad2021} or chemical gradients. 
%MST do we have references for any of these? --> (A) Fixed, some of them have already referred in the introduction
For the case in which the organism can sense a single lab frame signal direction in this way we found that a combination of these two signals (hydrodynamic forces and signal orientation), along with the memory of two recent actions, can yield the same qualitative level of performance as when using lab-frame information. Finally, we also investigate the effect of learning for different swimming modes, i.e. pusher, puller, and neutral. When given the same set of signals, pushers show the best performance, above neutral swimmers, with pullers performing the worst. 
%%MST maybe we tighten this up if we get some statistics on the sensors?
%We speculate that this might be connected with the clarity\JJM{What is the meaning of ``clarity'' here?} of the hydrodynamic signal on the sensors for different swimmers.\AOM{I think this sentence is added by MT or sensei. I presume he meant the ``comprehensibility'' in sensing stresses of each type.}
\section{Simulation methods}
\subsection{System of interest}
We consider a swimmer navigating through a Newtonian fluid with an imposed zig-zag shear flow. 
%MST I removed the small Re restriction, which we don't actually have. Also, we aren't actually at very small Re, so better not to claim it!
The coupled dynamics of the swimmer and the fluid are evaluated by solving a modified Navier-Stokes equation, which accounts for the fluid-particle interaction using the smoothed profile (SP) method \cite{Yamamoto2021}, together with the Newton-Euler equations for the rigid-body dynamics. The squirmer is assumed to be able to perceive information from its environment and perform actions accordingly. These actions are determined by a weighted neural network, trained using a Deep Q-Learning algorithm, to draw actions that lead to the highest accumulated reward over a given time interval.
%The instantaneous chosen actions at the beginning of the learning procedure, before 
%a good policy has been developed, are basically random. However, the choice of action %towards the end of the simulation improve markedly. Here it is still not a truly optimal %policy. Indeed its likely that this isn't achievable within our approach. Below we provide %details of the swimmer model, the fluid solver, and the learning scheme.

\subsection{The squirmer model}
Here, we consider the ``squirmer'' model to represent swimmers as self-propelled spherical particles with a modified stick-boundary condition \cite{Lighthill1952,blake_1971}. Originally, this model was proposed to describe the dynamics of ciliated micro-organisms, where the dynamics of the swimmers is driven by 
%radial, tangential, and azimuthal velocities 
the fluid flows generated at their surface. The expression for this surface velocity is given as an expansion in terms of Legendre polynomials. For simplicity, the radial and azimuthal components are usually neglected, and the expansion is usually truncated to neglect modes higher than second order \cite{Brenner1983}. Thus, the slip velocity of the swimmer at a given point on its surface is characterised by spherical polar variables $(\vartheta,\varphi)$, with $\vartheta=0$ corresponding to the swimming direction, according to
\begin{equation}
    \textbf{u}^s(\vartheta,\varphi) = B_1\left(\sin\vartheta + \frac{\alpha}{2}\sin2\vartheta\right) \bm{\hat{\vartheta}}
    \label{squirmeru}
\end{equation}
where $\hat{\bm{\vartheta}}$ is the tangential unit vector in the $\vartheta$ direction, $\vartheta = \cos(\hat{\bm{r}}\cdot\hat{\bm{e}})$ the polar angle, with $\hat{\bm{e}}$ the swimming axis, 
%MST see earlier comment - surely v here? -->(A) I use \hat{n}_e for the eye direction. So, it's a right notation here.
and $\hat{\bm{r}}$ a unit vector pointing from the center of the squirmer to the point $(\vartheta,\varphi)$ on the surface. The coefficient $B_1$ is the amplitude of the first squirming mode, which determines the steady-state swimming velocity of the squirmer $U= \frac23B_1$, with $\alpha = B_2/B_1$ characterizing the type of flow field: for negative (positive) $\alpha$, the squirmer is a pusher (puller), e.g. E. \textit{Coli} and C. \textit{reinhardtii}, respectively. 
%MST check correct taxonomic nomenclature - are both genus and species italicised? You must either shorten Chlamydomonas to C. or expand E. to Escheria (check my spelling here!) for consistency. --> (A) fixed
The first mode $B_1$ relates to the hydrodynamic source dipole, with a decay in the velocity field proportional to  $1/r^3$, while that of the second mode is related with a force dipole, which decays as $1/r^2$. For a neutral squirmer (e.g., \textit{Paramecium}), $\alpha = 0$, the first mode dominates over $B_2$ and the velocity field decays as $1/r^3$.

\subsection{The smoothed profile method}
To solve for the coupled fluid and particle dynamics, we solve the equations of motion for both the viscous host fluid and the squirmer using the smoothed profile (SP) method \cite{Yamamoto2021}. The evolution of the particle obeys the Newton-Euler equations:
\begin{align}
\dot{\bm{R}}_i &= \bm{V}_i, &\dot{\bm{Q}}_i &= \mathrm{skew}(\bm{\Omega}_i)\cdot\bm{Q}_i,\notag\\
M_p\dot{\bm{V}}_i&= \bm{F}^H_i + \bm{F}^C_i + \bm{F}^{ext}_i, &\bm{I}_p\cdot{\dot{\bm\Omega}}_i&= \bm{N}^H_i + \bm{N}^{ext}_i \label{eq2}
\end{align}
where $i$ is the particle index, $\bm{R}_i$ and $\bm{V}_i$ are the particle center-of-mass position and velocity, respectively, $\bm{Q}_i$ is the orientation matrix, and $\bm{\Omega}_i$ the angular velocity. 
The skew-symmetric matrix is defined such that $\mathrm{skew}(\bm{\Omega}_i)\cdot \bm{x} = \bm{\Omega}_i\times \bm{x}$ ($\forall \bm{x}\in \mathbb{R}^3$),
\begin{equation}\label{eq3}
    \mathrm{skew}(\bm{\Omega}_i) = \begin{pmatrix} 
0 & -\Omega^z_i & \Omega^y_i \\ 
\Omega^z_i & 0 & -\Omega^x_i \\ 
-\Omega^y_i & \Omega^x_i & 0  
\end{pmatrix} 
\end{equation}

The forces exerted on the particle, appearing in the rhs of \eqref{eq2}, include the hydrodynamic forces $\bm{F}^H$, the direct particle-particle forces $\bm{F}^C$, given by a truncated Lennard-Jones potential with powers $36-18$ 
%Wow, those are high exponents. This begs the question why the usual 12-6 isn't enough. Clarify? --> (A,J) the reason is you want it to be as hard-sphere-like as possible. So this 36-18 is a proper one, it's usually the one used in previous Kapsel studies, where the attractive interaction is always turned off.
(to prevent particle overlaps), and external forces $\bm{F}^{ext}$, e.g. gravity. Likewise, the torques are decomposed into hydrodynamic $\bm{N}^H$ and external $\bm{N}^{ext}$ contributions, where we have neglected inter-particle torques. The forces and torques are evaluated assuming momentum conservation to ensure a consistent coupling between the host fluid and the particles. The time evolution of the host fluid is determined by the Navier-Strokes equation, together with the incompressibility condition:
\begin{align}
    \nabla\cdot\bm{u}_f &= 0 \label{eq4}\\
    \rho_f(\partial_t + \bm{u}_f\cdot\nabla)\bm{u}_f &= \nabla\cdot\bm{\sigma}_f + \rho_f\bm{f}\label{eq5}\\
   \bm{\sigma}_f &= -p\bm{I} + \eta_f[\nabla\bm{u}_f+(\nabla\bm{u}_f)^T]\label{eq6}
\end{align}
where $\rho_f$ is the fluid mass density, $\bm{u}_f$ the fluid velocity field, $\eta_f$ the shear viscosity, $\bm{\sigma}_f$ the stress tensor, and $\bm{f}$ an external body force. 

When applying the SP method, the sharp interface between the rigid particle and the fluid domains is replaced by an interfacial region with a finite width $\xi$, with both regions characterized by a smooth and continuous function $\phi$. This function returns a value of 0 for the fluid domain and 1 for the solid domain. The total velocity field, $\bm{u}$, can then be written as
\begin{equation}\label{eq7}
   \bm{u} = (1 - \phi)\bm{u}_f + \phi\bm{u}_p
\end{equation}
The first term on the RHS of \eqref{eq7} represents the contribution from the host fluid, while the second term is from the rigid-body motion. Then, we can consider the system as a single-component fluid, and write down a modified Navier-Stokes equations, similar to \eqref{eq5}, but in terms of the total velocity $\bm{u}$, as
\begin{equation}
   \rho_f(\partial_t + \bm{u}\cdot\nabla)\bm{u} = \nabla\cdot\bm{\sigma}_f + \rho_f(\phi\bm{f}_p + \phi\bm{f}_{sq} + \bm{f}_{shear})\label{eq8}
\end{equation}
where $\phi\bm{f}_p$ is the force density field required to maintain the rigidity of the particle, $\phi\bm{f}_{sq}$ is the force density required to maintain the squirming motion, and $\bm{f}_{shear}(\bm{x},t)$ is an external force required to maintain the following zig-zag velocity profile \cite{Iwashita2009}:
\begin{equation}\label{eq9}
   v_x(y) = 
\begin{cases}
    \dot{\gamma}(-y - L_y/2),& -L_y/2 < y\leq -L_y/4\\
    \dot{\gamma}y,& -L_y/4 < y\leq \phantom{-}L_y/4\\
    \dot{\gamma}(-y + L_y/2),  & \phantom{-}L_y/4 < y\leq \phantom{-}L_y/2
\end{cases}
\end{equation}
where $\dot{\gamma}$ is the shear rate, $y$ the distance in the velocity-gradient direction, and $L_y$ the height of the three-dimensional rectangular simulation box, of dimensions $(L_x,L_y,L_z)$. We numerically solve the equations of motion using a fractional step procedure. First, the  total velocity field is updated by solving for the advection and hydrodynamics stress contributions in the Navier-Stokes equation. Simultaneously, the particle positions and orientations are propagated forward in time. Second, we evaluate the momentum exchange over the particle domain, and use it to compute the hydrodynamic contributions to the forces (torques) exerted on the particles. Third, the updated forces and torques are used to update the particle velocities, in such a way that the squirming boundary condition is maintained (through $\phi\bm{f}_{sq}$). Finally, the rigidity constraint ($\phi\bm{f}_p$) is computed, in such a way that the momentum conservation is guaranteed, and used to update the total velocity field (together with the shear-flow constraint $\bm{f}_{shear}$). A detailed discussions of this procedure can be found in our earlier work\cite{Iwashita2009,John2013s,Yamamoto2021}.

\subsection{Deep reinforcement learning}
We employ a Reinforcement Learning (RL) framework \cite{Sutton2018} to obtain optimal policies for the prescribed swimming tasks. 
%MST is this the right place to introduce this fact? --> (A) I am not so sure where else it can be fit. I think giving a big picture including what the method has achieved at the beginning is quite reasonable. 
This involves training a neural network to select actions that generate a high reward. RL has proven itself to be a powerful tool for finding flow control and navigation strategies \cite{gazzola2016,Colabrese2017,Reddy2018}. In RL, an agent (here the swimming particle), uses information received from its environment to define its current \textit{state}, which it uses to determine its next \textit{action}, resulting in a corresponding \textit{reward} (assumed to be a real number) for this action. This type of agent-environment interaction allows one to control the agent decisions, in order to maximize the long-run accumulated reward without prior knowledge of the dynamics of the system. In this study, we adopt a Deep Q-learning strategy, combined with prioritized experience replay and n-step learning \cite{vanhasselt2015,schaul2016pri,Sutton1988}, as the search tool for finding optimal navigation strategies for a given task. 

\begin{figure}[ht!]
    \centering
    \includegraphics[width=\columnwidth]{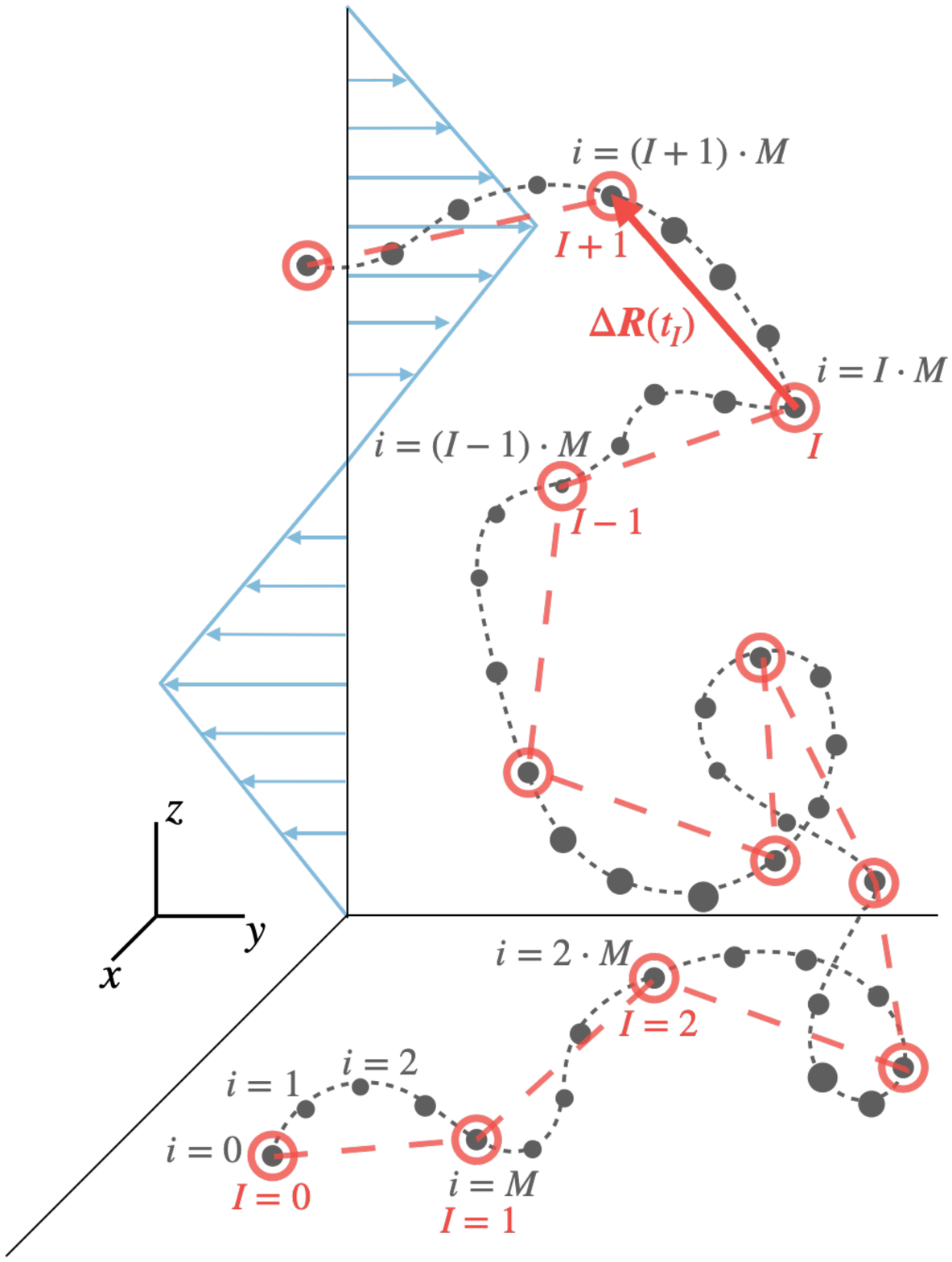}
   \caption{A schematic illustration of a particle learning to navigate a zig-zag shear flow. The particle trajectory over a learning episode is illustrated as the (black) short-dashed line. The position of the swimmer at the discrete simulation time steps $i$ is marked with (black) filled circles. This trajectory is \textit{coarse-grained} to define the action trajectory, illustrated as the (red) long-dashed line. This action trajectory consists of action segments, composed by taking every $M$ simulation steps, marked with (red) open circles. The swimmers chooses an action $a_I$ at the start of each action segment $I$, which it follows until the start at the next action segment ($I+1$).
}
 %MST notation - important - make the r, R notation from this fig consistent with that used in the text. I prefer use \Delta R in place of r_total. The rewards, e.g. \Delta z as shown on the results figs, are further inconsistent with the text where you write \Delta R,z etc. This needs sorting! --> (A) In the text, the r_t^d where d is the reward direction is used. So, I changed that in the cartoon.
 %e is the eye direction, isn't it? Surely v is better here? Or show both. --> (A) \hat{n}_e is the eye direction \hat{e} is the swimming direction
 %What are the red dots on the red trajectory - they are not explained in the caption or elsewhere. What is an "individual time step"? If its defined in the text say so (maybe also say in the caption if the particle moves <<, ~ or >> its size in this time step). --> (A) just gives a definition of the red dots, John actually invented this to show the location of the swimmer at each time step in the trajectory.
    \label{fig0}
\end{figure}

The swimmer is trained by maximizing the expected reward over a fixed time-interval, called an episode (see Fig.\ref{fig0}). Episodes are discretised into $N_s$ action segments of $M$ simulation steps each, such that $T_{\mathrm{episode}} = N_s\cdot~\Delta T$, with $\Delta T = M\cdot \Delta t$ the time-duration of the action segment ($\Delta t$ the simulation time step). Let $s_I$ be the state of the swimmer at the beginning of action segment $I$, which corresponds to simulation time step $i=I\cdot M$ and time $T_I = I\cdot \Delta T = (I\cdot M)\cdot \Delta t \equiv t_{i=I\cdot M}$. The swimmer uses a policy function $\pi$, which maps states to actions, in order to choose its next action $a_I$, which it will follow for the duration of the action segment (i.e., the next $M$ simulation steps), after which it will be in a new state $s^\prime_I = s_{I+1}$. The swimmer is then assigned a reward $r_I$, which depends on its state at the endpoints, $s_I$ and $s_I^\prime$. This gained (action-reward) \textit{experience} is written in tuple form as $(s_I, a_I, s^\prime_I, r_I)$. For the purposes of the learning, the state of the system at intermediate times between the start of subsequent actions steps $I$ and $I+1$, i.e., $T_I = t_{I\cdot M}< t < t_{(I+1)\cdot M} = T_{I+1}$,  will be irrelevant. Finally, at the end of the episode, the total reward is evaluated by accumulating each of the individual action rewards acquired during the trajectory, $r = \sum_{I=0}^{N_s-1} r_I$. 

To compute the (optimal) policy, define the action-value function $Q_\pi$, for a given policy $\pi$, as $Q_\pi(s_I,a_I) = r_{I} + \gamma r_{I+1} + \gamma^2 r_{I+2} + \ldots$, with $\gamma\in[0,1)$ a discount factor. This $Q$ function gives the expected accumulated reward for adopting action $a_I$ during step $I$ (starting from state $s_I$), expressed as the reward for action step $I$, plus the (discounted) rewards at each subsequent step ($0< I < N_s$). The optimal policy function $\pi^*$, whose mapping of states and actions maximizes the long-time reward must satisfy the Bellman equation $Q_{\pi^\star}(s_I,a_I) = r_{I} + \gamma \max_a Q_{\pi^\star}(s_{I+1},a)$ \cite{Sutton2018}. To learn this optimal policy the $Q$ value function is represented by a neural network and trained in an episode-based fashion, over $N_{\mathrm{episode}}$ episodes, with each episode consisting of $N_s$ actions steps, of $M$ simulation steps each.

Throughout the training phase, at the beginning of each episode the position and orientation of the swimmer is randomly set, and the swimmer is allowed to navigate for the time duration $T_{\mathrm{episode}}$ ($N_s$ action segments). At each action step $I$, the (current) best action is that which maximises the (current) value function, i.e., $a_I = \mathrm{argmax}_{a}Q_\pi(s_I, a)$.  In order to reduce the bias in the training, a batch of stored experiences of size of $N_{b}$ is drawn from a ``replay memory'' buffer (size $N_{P_{max}}$) at the end of each action step. The replay memory buffer will store the experiences gathered over many action steps and episodes. Each element of this batch of experiences consists of information on the environmental signals, the actions taken, and the immediate rewards, i.e. $(s_I, a_I, s^\prime_I, r_I)$. The drawn batch is then used to adjust the weights of $Q$, according to the following rule
\begin{align}
    Q(s_I,a_I) &\leftarrow Q(s_I,a_I) \label{eq10}\\
    &+ \alpha[r_{I} + \gamma \max_a Q(s_{I+1},a) - Q(s_I,a_I)]\notag
\end{align}
where $\alpha$ is the learning rate. The $Q$ network is trained against the following loss function (with $\bm{\theta}$ the network weights),
\begin{align}
    L_I(s_I, a_I; \bm{\theta}_I) = (Y_I - Q(s_I,a_I;\bm{\theta}_I))^2\label{LN1}
\end{align} 
where $Y_I$ is the ``target'' at learning step $I$, defined as
\begin{align}
    Y_I = r_I + \gamma \max_a Q(s_{I+1},a;\bm{\theta}^-_I)\label{LN2}
\end{align}
with $\bm{\theta}^-_I$ a set of target network parameters which are synchronized with the ``prediction'' Q-network parameters (the ones begin optimized for) $\bm{\theta}_I^- = \bm{\theta}_I$ every $C$ steps, and otherwise held fixed between individual $\bm{\theta}$ updates. This can be understood as a loss function that depends on two identical Q-networks, a prediction network and a target network, but is only trained on the former. The gradient with respect to the weights $\bm{\theta}$ can then be written as (writing only the $\bm{\theta}$ dependence): 
\begin{align}
   \nabla_{\bm{\theta}_I}L(\bm{\theta}_I)= [Y_I - Q(\bm{\theta}_I)]\nabla_{\bm{\theta}_t}Q(\bm{\theta}_I)\label{LN3}
\end{align}
We use the ADAM optimizer \cite{adsm2014} to minimize the loss function. A detailed description of the deep Q-learning framework we have used can be found in ref.\cite{vanhasselt2015}.

%in one of the three characteristic flow directions $x$ (flow), $y$ (shear gradient) or $z$ (vorticity), depending on the assigned task in each case.
%MST this statement is incorrect for rms (unsigned) rewards, so I removed it. 
%This total reward is computed by accumulating the individual rewards at each time step.  
%MST We have a problem: this is inconsistent with the definition of these variables as being per timestep that appears a few lines later!
%These displacements are written $\Delta x$, $\Delta y$ and $\Delta z$ respectively.
% is this correct or is the reward the displacement divided by total time steps? I think the way we normalise our reward means it doesn't now matter, right? --> (A) This is correct, the "Total reward" is the total displacement divided by the maximum possible displacement in the direction. So, should I uncomment the below sentence?
%Unless stated otherwise, these single-step rewards are always given by the signed displacement value over the time step $\Delta t$. 
%MST I just commented this definition out. Is it needed elsewhere? $\Delta R_\alpha(t) = R_\alpha(t) - R_\alpha(t-\Delta t)$. 

In order to maximize exploration of the phase space, particularly at early stages of the learning, we have adopted an $\epsilon$-greedy selection scheme. Thus, the chosen policy $a_I$ is allowed to deviate from the optimal policy. That is, the optimal policy determined from the action-value function $Q$ is used with probability $1-\epsilon$, otherwise the action is randomly drawn from the action space with probability $\epsilon$. This greedy parameter is exponentially decaying in time, starting from $\epsilon=1$, until it reaches a value of $\epsilon=0.015$. The reason for decaying the greedy parameter is to prevent the swimmer from prematurely narrowing down the state-action space. During the early episodes of the training, the policy is very far from its optimum, thus, the swimmer needs to explore the state-action space as much as possible to gain experience interacting with the environment. Thus, it is better to draw random actions, rather than sticking to a specific policy. However, after a suitable training period, the policy is expected to improve, and the swimmer should now favor the trained policy, rather than a randomly chosen action. Note, that we always allow for a small probability $\epsilon$ for selecting a random action, even though the policy is expected to converge to the optimal one, since we aim to provide some room for possible improvements of the current best policy.

Finally, we will consider two forms of reward, signed rewards, in which the reward for segment $I$ is computed from the displacement of the swimmer $\Delta \bm{R}(T_I) = \bm{R}(T_{I+1}) - \bm{R}(T_I)$, and unsigned rewards, computed from the absolute value of the displacements. Furthermore, since we consider the three distinct tasks of swimming in the shear-flow ($x$) directions, shear-gradient ($y$), and vorticity ($z$) directions, the rewards are given by $r_I = \vu{e}^\alpha\cdot\Delta\bm{R}= \Delta R^\alpha(T_I)$, in the signed case, and $r_I = \abs{\Delta R^\alpha (T_I)}$, in the unsigned case ($\vu{e}^{\alpha}$ the unit basis vectors in the lab frame, $\alpha=x,y,z$). The latter type of reward may be a natural choice for organisms in which there is no preferential direction of motion, but where moving to a new location may be advantageous. An example of this could be the swimming of the marine bacterium, \textit{V. alginolyticus}, whose swimming pattern is a cycle of forward and backward swimming, together with turns to change its direction of motion\cite{Son2013}.  Furthermore, we discretise the action space, and define an action to be an external torque $\bm{N}^{ext} = H\hat{\bm{n}}$ that the swimmer can activate, with $H$ 
%MST You can't just say this without explaining your dimensionless units. I imagine these all need carefully defining. These are probably best introduced, and carefully defined, when you set up the physics of the simulation procedure, but it can be anywhere - as long as its before you write this!
the magnitude of the torque, and $\hat{\bm{n}} = \bm{m}/|\bm{m}|$ a unit vector with $m^\alpha = -1, 0, 1$.
Thus, the size of the action space is given by the $3^3 = 27$ possible rotation axes.
%MST I now see rewards are per time step. You say they are signed. Do we ever say "otherwise"? We report total rewards in the figures, right? How do we relate unsigned rewards per time step and overall? Please discuss/email with me Aom. --> (A) I think the one you commented out in 181th line states this. Should I uncomment that line?

%MST I don't knwo what the following four sentances mean, so neither will a reader. More careful discssion required --> (A) it's the explanation of the memory storage I mentioned to you, I rephrased it.

\begin{figure*}[htp!]
  \centering
  \includegraphics[width=\textwidth,height=5cm]{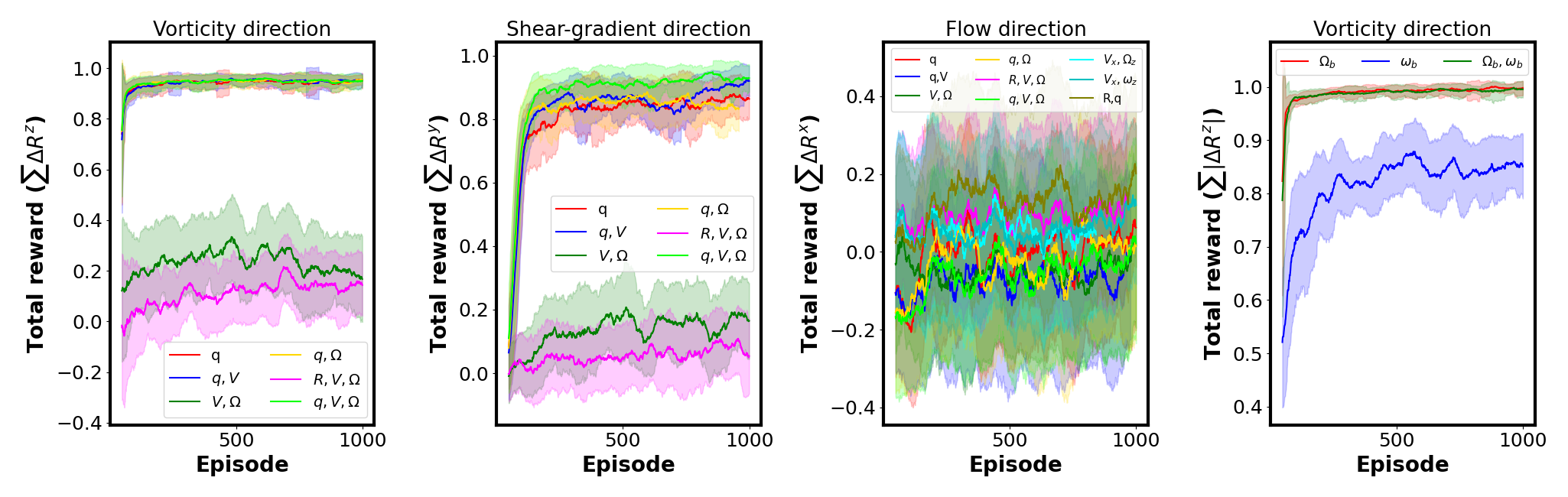}
  %MST typo on the fig labels "swim in across the shear gradient"-> "swim in the shear gradient". 
  \caption{\label{fig1}Learning performance for different tasks/rewards as measured by the rolling average of the total reward, normalised by the maximum possible displacements per episode. From left to right: Learning to swim in the vorticity ($z$), shear-gradient ($y$), and shear-flow ($x$) directions, using lab-frame information and signed rewards, and finally, learning to swim in the vorticity direction using body-frame information and unsigned rewards. For each task, we train our swimmer using different sets of signals, which can include the orientation quaternion (q), translational velocity ($\bm{V}$), rotational velocity ($\bm{\Omega}$), position ($\bm{R}$), and background flow vorticity ($\bm{\omega}$), as specified in the plot legend. The two previous actions ($\textbf{a}$) were included in all cases (not labelled). The subscript $b$ in the legend of the last panel denotes signal variables projected into the body-frame. An averaging window size of 50 episodes was used. The shaded area represents the standard error in the mean. The signed rewards are defined as $\sum_{I = 0}^{N_s -1}\Delta R^\alpha(T_I) = \sum_{I = 0}^{N_s -1} (R^\alpha(T_{I+1})-R^\alpha(T_I))$ where $I$ is the action/learning step and $N_s$ the total number of action segments per episode. Likewise the unsigned rewards given as
$\sum_{I = 0}^{N_s -1}|\Delta R^z(T_I)| = \sum_{I = 0}^{N_s -1}|R^z(T_{I+1})-R^z(T_I)|$.}
\end{figure*}

\subsection{The system parameters}%MST I suggest you move this entire section forward to solve the H=400 issue. --> (A) a bit troublesome in the sense since some of the parameter has just been described after the H = 400. So, I just left the description of the H above and put the number below.
Throughout this work we present our results in simulation units, using as basic units of length, density, and viscosity the grid spacing $\Delta=1$, fluid density $\rho_f=1$ and viscosity $\eta=1$. The units of time and mass are $\rho_f(\Delta)^2/\eta = 1$ and $\rho_f \Delta^3 = 1$, respectively.
The radius of the spherical swimmer is $\sigma = 5\Delta$, 
%MST Define the simulation timestep Delta t in these units (I make it Delta t=(sigma/10)^2 rho_f/eta=1 but please check!), and hence the dimensionless runtime T in these units (where Delta t=1) equivalent to your N and N_step - see comment below. --> (J?) --> I've just add the definition of the time step above.
%%% JJM see above.
the size of our rectangular simulation box is $32\Delta \times 64\Delta \times 32\Delta$, with full periodic boundary conditions along all dimensions. Other parameters used in the SP simulator are the particle-fluid interface thickness $\xi = 2\Delta$, the particle density $\rho_p = \rho_f$ and the magnitude of the external torque $H = 400 \eta^2 \Delta/\rho_f$. The applied shear rate is $\dot{\gamma} = 0.04\eta /(\rho_f \Delta^2)$, which corresponds to a Reynolds number of $Re \approx 1$. For most of the cases presented here, and unless stated otherwise, the swimmer is set to be a puller with $\alpha = 2$ and $B_1= 0.1\eta/(\rho_f\Delta)$, corresponding to a particle Reynolds number of $Re \approx 6 \times 10^{-2}$,  comparable to that of E. \textit{coli} in water \cite{Purcell1977}.

To further characterize our system we introduce the following three dimensionless $\psi$ parameters,
\begin{align}
    \psi_1 &= %\frac{V_{\mathrm{swim}}}{V_{\mathrm{shear}}} = 
    \frac{\frac23 B_1}{\frac{\dot{\gamma}}2 L_y} \\
    \psi_2 &= %\frac{\delta_{\mathrm{ext}}}{\delta_{\mathrm{shear}}} = 
    \frac{\frac{H}{\pi \sigma^3 \eta}}{{\dot{\gamma}}/2} \\
    \psi_3 &= %\frac{V_{\mathrm{swim}} N_\mathrm{step}}{L_y} = 
    \frac{\frac23 B_1 T_{\mathrm{episode}}}{L_y}
\end{align} %MST I commented out these statements because its easier like this I think The Delta t factors in the second cancel!
%MST I don't like the notation N_step. You use just N in Fig 1. I prefer something like T, the run time in Delta t units. Please make consistent throughout. Note I changed N_step to T in psi_3. --> (J?) --> I fixed the description but needs your help to reread from L286 - 304 again.
These measure the strength of the swimming in three natural ways: $\psi_1$ is the ratio of the baseline swimmer speed to the maximum shear flow speed (twice the typical shear flow speed); $\psi_2$ is the ratio of the active rotation rate of the swimmer to that induced by the shear flow;  $\psi_3$ is the ratio of the maximum total active displacement of a swimmer over one episode, duration $T_{\mathrm{episode}}$, 
%MST note I use T here so you'll need to fix this if you decide to use something else.
to the largest system size $L_y$. Unless otherwise specified our system parameters correspond to $\psi_1\simeq 5\times 10^{-2}$, meaning the swimmer moves slowly compared to the fluid speed; $\psi_{2}\simeq 6$, meaning that the swimmer can actively rotate faster than the rotation induced by the shear flow, necessary for it to have meaningful control of its orientation;
%MST give the real space and angular displacements per time step here. (A) --> Added in the below paragraph.
and $\psi_3 \simeq 20$, meaning the swimmer can explore regions with different shear gradients. The characteristic angular rotation (per epoch) caused by the external torque corresponds to $\frac{H}{\pi \sigma^3 \eta}T_{\mathrm{episode}}\simeq 180 \,\mathrm{rad}$, meaning that the agent can fully rotate many times.  For the learning parameters, we use a discount rate $\gamma = 0.93$, learning rate $\alpha = 0.00025$, and batch size $N_b = 128$, with a replay memory size of $N_{P_{max}} = 100000$ and greedy parameter ($\epsilon$)
%MST epsilon appears undefined here, as far as I can tell? --> (A) it has been mentioned in L258
 decay rate of $k = 0.981$. 
 %MST whats this? define it if not previously. --> (A) same as the above comment.
 The neural network consists of 1 input layer with a number of neurons equal to the number of state-defining variables,
 %MST say "typically...'' and give a number, perhaps? --> (A) but 27 is not typical... it only applies for this case where we have 27 actions that can be chosen. 
 with 3 hidden layers of 100 neurons each, and 1 output layer with 27 neurons, corresponding to the size of action space. Finally, a learning episode consists of  $N_s=2\times 10^3$ action steps of $M=10$ simulation steps each. The precise numerical  values for all our parameters can be found in the SI.
 
 %fwith each 

\section{Results and discussion}
One of the main challenges of RL is defining an appropriate state. Gunnarson \textit{et al.} \cite{Gunnarson2021} have shown that in unsteady two-dimensional flow fields, different sets of environmental cues lead to significantly different levels of performance for a given task.
%MST need we say this here? --> (J?) --> (A) I also dont think we need to say this . . .
%First, we train the swimmer to navigate through an applied shear-flow, with the goal of maximizing the migration distance in the vorticity direction (perpendicular to the shear plane). 
Here, to define the state we use combinations of the swimmer's lab-frame configuration, i.e., position $\bm{R}$, translational velocity $\bm{V}$, rotational velocity $\bm{\Omega}$, and rotation quaternion $q$, together with the background flow information, in particular the flow vorticity $\bm{\omega}=\bm{\nabla}\times\bm{u}$. Taken together, these sets of signals provide a complete specification of the swimmers current configuration.
%MST right? don't we need to define its initial orientation, from which it is rotated ? --> (A) line 174 already mentioned that the initial orientation is random, so . . .
We also include the swimmer's previous two actions, denoted \textbf{a} in all examples, with the goal of improving the convergence of the policy. Such limited memory may be accessible, even in micro-organisms \cite{YANG2020417}. Fig.~\ref{fig1} shows the rolling average of the normalised total rewards for the different swimming tasks. The normalization constant is defined as the maximum possible displacements per episode, calculated as $(2/3B_1) T_{\mathrm{episode}}$. 

For the tasks of swimming in the vorticity and shear-gradient directions, it is sufficient for the swimmer to receive orientation information, via the rotation quaternion ${q}$, along with the memory of the last two taken actions, in order to develop an efficient policy.
%MST did you try just q? --> (A) Yes, and the stability of the trained policy is slightly inferior, but that trial used the previous set of the system parameters. 
We  see that the signal variable combinations that include the orientation $q$ can approach an optimal policy, while those that don't show  inferior performance. In contrast, for the task of swimming in the flow direction, the swimmer was unable to reliably perform the task. Even the signals including the orientation information are no longer sufficient to help the swimmer to locate the flow speed maxima. 
%%MST what does the following refer to? Is any data shown. What was our reason for supplying exactly this "help", and not something else? --> (A) to see if these information is given to the swimmer, will the performance become better. The first two panels are plotted with 6 sets of inputs but the 3rd one got 9, the last three are the additional trials information. Apparently, giving information like the instantaneous location and orientation does improve the performance a bit but the swimmer still has the hard time to align itself to the maxima. --> should I uncomment the below paragraph?
%To help with this learning, we have also provided additional signals to the swimmer, i.e. the $x$ component of the velocity, the $z$ component of the angular velocity, and the $z$ component of the background flow's vorticity, and even the position and orientation $\{\bm{R},q\}$. The later signal seems to outperform the others, but this performance is still poor compared to that achieved for the other two tasks.  While the swimmer only needs to put itself in a configuration that allows it to be advected by the flow, it is unable to learn such a strategy. 
This may be because the swimmer has no direct memory of its previous configuration. A detailed quantitative analysis of the performance obtained for this task can be found in the SI. The last panel of Fig.~\ref{fig1} shows the results obtained when using unsigned rewards, defined as $r_I = |\Delta R^z(T_I)| = |R^z(T_{I+1}) - R^z(T_I)|$, for the task of swimming in the vorticity ($z$) direction. The signals used for this set of simulations are different (compared to learning with signed rewards),
%MST why? --> (A) because it senses the body frame rotational velocity or the flow vorticity rather than q,r,v ...
as the swimmer only needs to perceive how it is orientated relative to the flow vorticity $\bm{\omega}$, as measured by the angular velocity in the body frame $\bm{\Omega}_b$, in order to develop an efficient policy.
%%MST this seems repeated? --> (A) The above one is the figure caption, should I uncomment the below paragraph? 
%We have also considered using the flow vorticity (in the body frame) directly as an input signal. From the plot, one can clearly see that the swimmer learns to align its swimming direction to the vorticity direction when provided with information of the rotational velocity. 
While the signal from the background vorticity also provides information about the orientation of the swimmer, the performance in this case is not as good as that obtained when using the swimmer's rotational velocity.
%MST any idea why?! (A) I think since we already projected that to the body frame, the rotational velocity provide more accurate information of how it align with the design plane?

\begin{figure}[htp]
    \centering
    \includegraphics[trim={0.25cm 0 0 1.0cm},clip,width=0.48\textwidth]{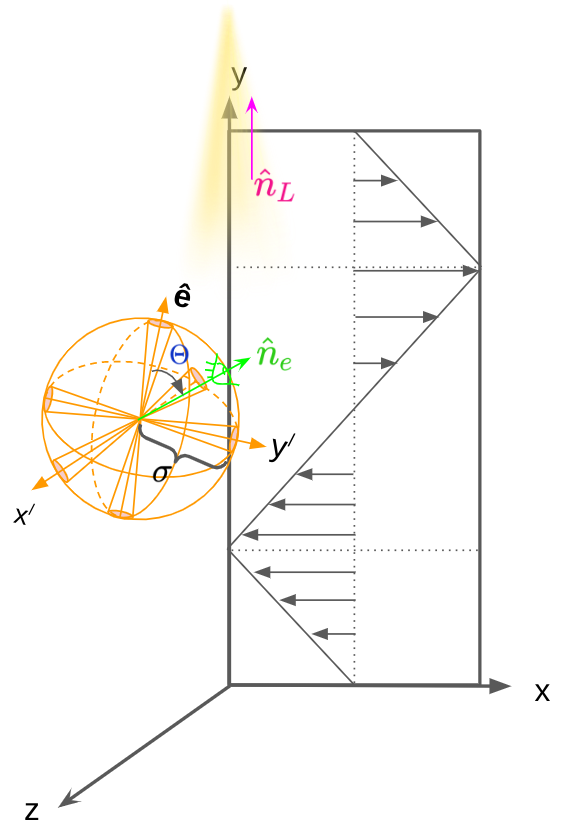}
    \caption{Schematic representation of a spherical microswimmer that measures (local) body-frame information while navigating a zig-zag shear flow. Here $x'$, $y'$, and $z'$ are the principle axes in the body reference frame, and $\hat{\bm{e}}=\hat{\bm{z}}$ is the swimming direction. We consider a swimmer with six sensors distributed across its surface, each one aligned with one of the principle axes. The stress sensors occupy spherical caps (highlighted) with an area of $\pi/2 \sigma^2(1-\cos{30})\approx0.21\sigma^2$ each. 
     %MST check you have been clear about exactly how the stress on these is calculated
     The swimmer is assumed to have a single sensor (eye) that can sense a signal (light), with a location specified by the unit vector $\hat{\bm{n}}_e$ (green arrow). The signal source can lie in any direction $\hat{\bm{n}}_L$ (purple arrow), but here we illustrate the case in which it is aligned with the shear-gradient direction (purple arrow).}
  \label{fig2}
\end{figure}
%MST check I converted all occurrances of global->lab frame and local -> body frame --> (A) fixed

%MST remove all references to "bug" and replace with micro-organism or model micro-organism or simply microswimmer. Sorry for using this colloquial terkm for ease of communication but its not really suitable for use in writing in a paper. --> (A) fixed
We have demonstrated the ability of idealised microswimmers to efficiently perform swimming tasks using lab frame information. 
%MST this feels clunky and feels poorly structured. Why did we just analyse the body frame vorticity signal response for (i) unsigned (ii) z rewards? --> (A) Fixed the sentence.
We have also shown that the body frame rotational velocity can be used to efficiently swim in the (unsigned) vorticity direction, since it can give a hint on the axis of rotation, which itself can be related to the alignment in the vorticity direction. However, active microorganisms in nature may not have such privileged information. Typically they can only obtain certain body-frame signals. A good example is that of copepods, 
which are able to sense the proximity of predators through the induced bending patterns of their setae, hair-like structures on the surface\cite{Kiorboe1999}. Recent work on RL for swimming in non-uniform flows has included local signals, e.g.local fluid strain rate and slip velocity, but the full hydrodynamic effects have not been taken into account\cite{Qiu2022_2,Qiu2022}. Furthermore, these studies have only considered the task of swimming against gravity. Given that swimmers can be expected to sense local surface stress signals, we train a swimmer to achieve a similar suite of swimming tasks as before (i.e., swimming in the shear-flow, shear-gradient, and vorticity directions), but using a set of physiologically reasonable body frame signals. First, the swimmer is assumed to be able to detect surface stresses, through the hydrodynamic forces exerted by the surrounding fluid on the swimmer, denoted as $\bm{\tau}_i$. Here, we assume that the spherical swimmer has six surface sensors ($0\le i < 6$), located on the antipodal points at the intersection of the three principle body-axes (see Fig.~\ref{fig2}). Each sensor is assigned  a surface area of $0.21 \sigma ^2$,  where $\sigma$ is the particle diameter. 
%MST is this the value you gave before? If not why not? --> (A) yes in the figure 3 caption L334
These sensors are given a finite size in order to average the surface stress derived from the numerical simulations. 
%MST is this the reason. Its a weak one. Why couldn't we do it at a point? Are we clear enough about how this is averaged? --> (J?) (A) I dont think we have mentioned the averaging process . . .
%JJM I would assume they would be normalized by area ? <f> = \int da f / \int da ?
Furthermore, we also assume that this swimmer has a sensor (eye) located on its surface, at an angle $\Theta$ from the swimming direction $\vu{e}$.
%MST you need to think of a new variable name for this as you already have theta for the surface coordinate variable. How about capital \Theta? need to fix everywhere, including maybe on the actual schematic figures! --> (A) to be fixed
This sensor can be considered to detect visual cues (i.e., light), via the signal $\hat{\bm{n}}_e \cdot \hat{\bm{n}}_L$. This might, therefore, serve as a model for a micro-organism capable of migrating towards/away from light sources.
%MST you just used another different notational convention for the eye direction n from the one in the schematic figure n_e. Fix to ensure consistency throughout. --> (A) I think I always use \hat{\bm{n}}_e for the eye direction, not sure which one you meant?
We consider cases in which the light can come from one of three directions: parallel to the flow direction ($x$), the shear-gradient direction ($y$), or the vorticity direction ($z$). In addition, we also consider the case in which the model micro-organism has the ability to sense the ``flashing'' of the light due to its relative reorientation. This flashing signal is encoded as $(\hat{\bm{n}}_e \times \bm{\Omega})\cdot\hat{\bm{n}}_L$. We utilize these as state-defining parameters and repeat the same Q-learning procedure used previously, in order to obtain the optimal policy for each of the three swimming tasks, i.e., swimming in the shear-flow, shear-gradient, and vorticity directions.

\begin{figure*}[ht!]
    \centering
    \includegraphics[trim={0.3cm 0 0 0.2cm},clip,width=0.9\textwidth]{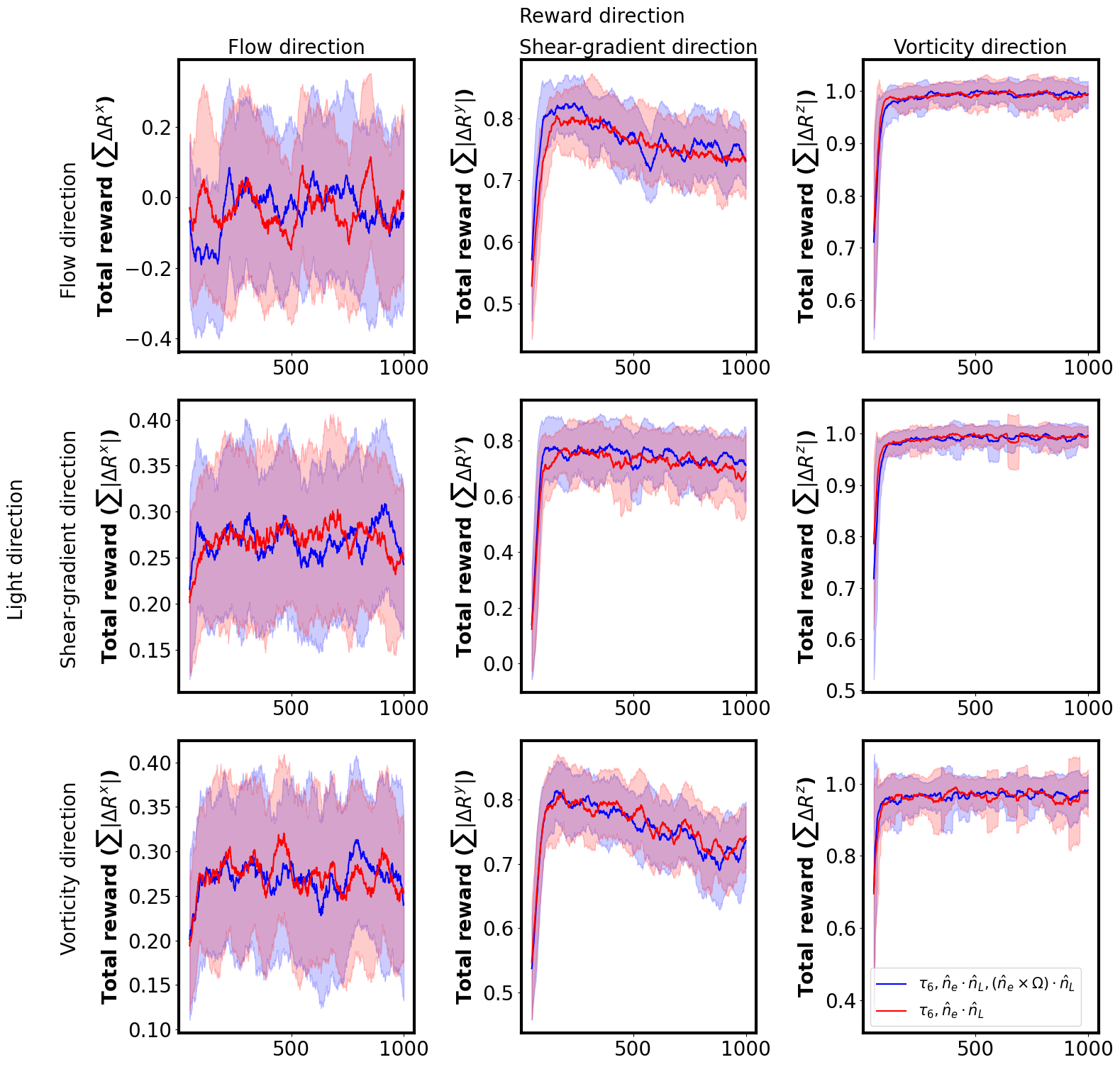}
    \caption{The rolling average of the normalized total rewards, with an averaging window of 50 episodes, for swimmers tasked with migrating in different directions $\alpha$, under light sources shining along direction $\beta$ ($\alpha,\beta = x,y,z$). Columns represents the reward direction $\alpha$, rows the signal direction $\beta$. The total rewards are defined either in terms of signed displacements $\sum_I R^\alpha(T_I)$ ($\alpha=\beta$), or unsigned displacements $\sum_I \abs{\Delta R^\alpha(T_I)}$ ($\alpha\ne \beta$). Here we consider two sets of input signals, $\{\bm{\tau}_i,\hat{\bm{n}}_e\cdot\hat{\bm{n}}_L,(\hat{\bm{n}}_e \times \bm{\Omega})\cdot\hat{\bm{n}}_L\}$ and $\{\bm{\tau}_i,\hat{\bm{n}}_e\cdot\hat{\bm{n}}_L\}$ with the eye's location at $\Theta = 30^\circ$.}
    \label{fig3}
\end{figure*}

\begin{figure*}[ht!]
    \centering
    \includegraphics[trim={0.3cm 0 0 0.2cm},clip,width=0.9\textwidth]{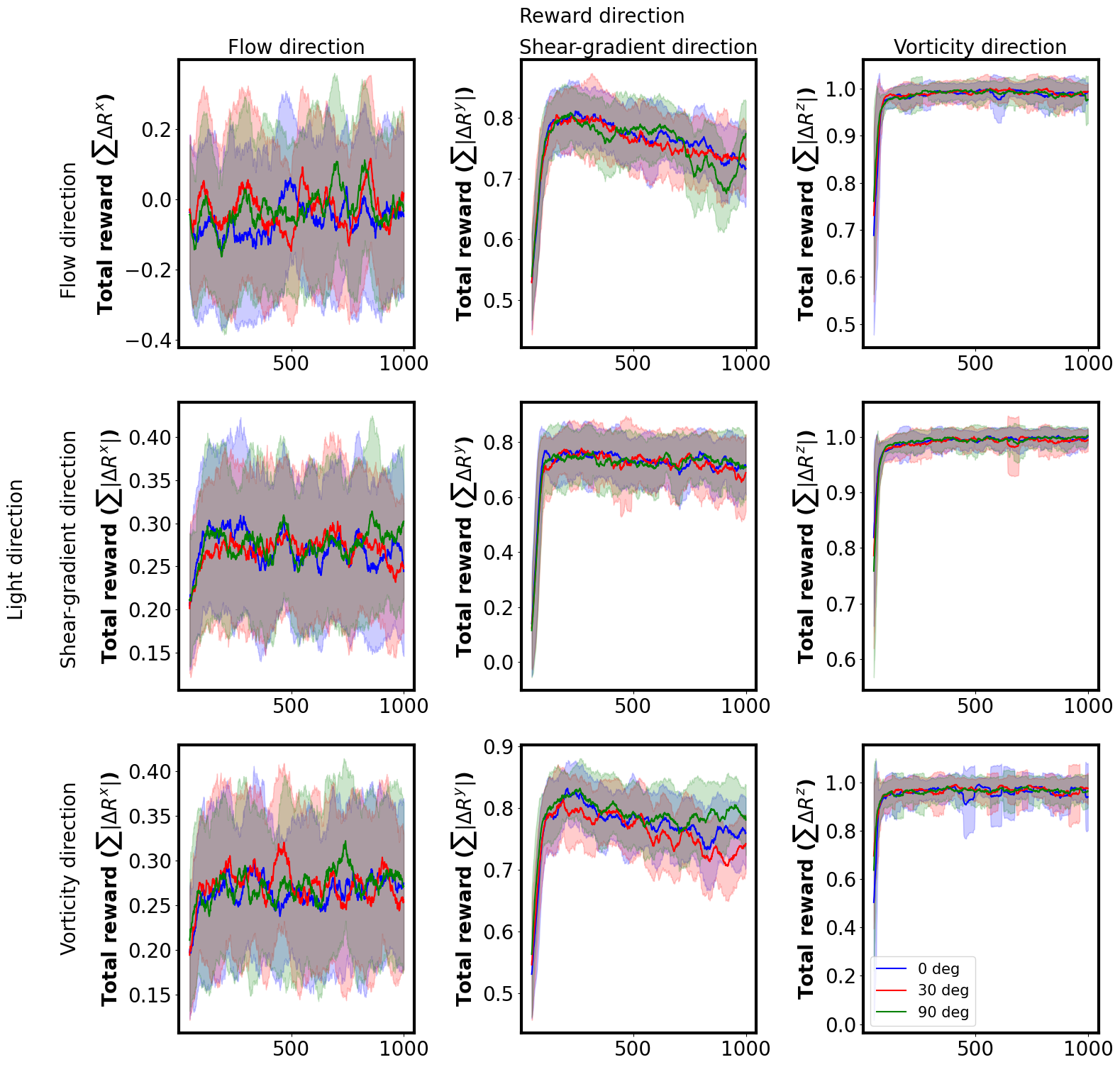}
    \caption{The rolling average of the normalized total rewards, with an averaging window of 50 episodes, for visually aware swimmers tasked with migrating in different directions $\alpha$, under light sources shining along direction $\beta$. We consider swimmers with eyes located at  $\Theta = 0^\circ, 30^\circ \text{and}\ 90^\circ$, using $\{\bm{\tau}_i,\hat{\bm{n}}_e\cdot\hat{\bm{n}}_L\}$ as the state defining variables.}
    \label{fig4}
\end{figure*}

\begin{figure*}[ht!]
    \centering
    \includegraphics[trim={0.3cm 0 0 0.2cm},clip,width=0.9\textwidth]{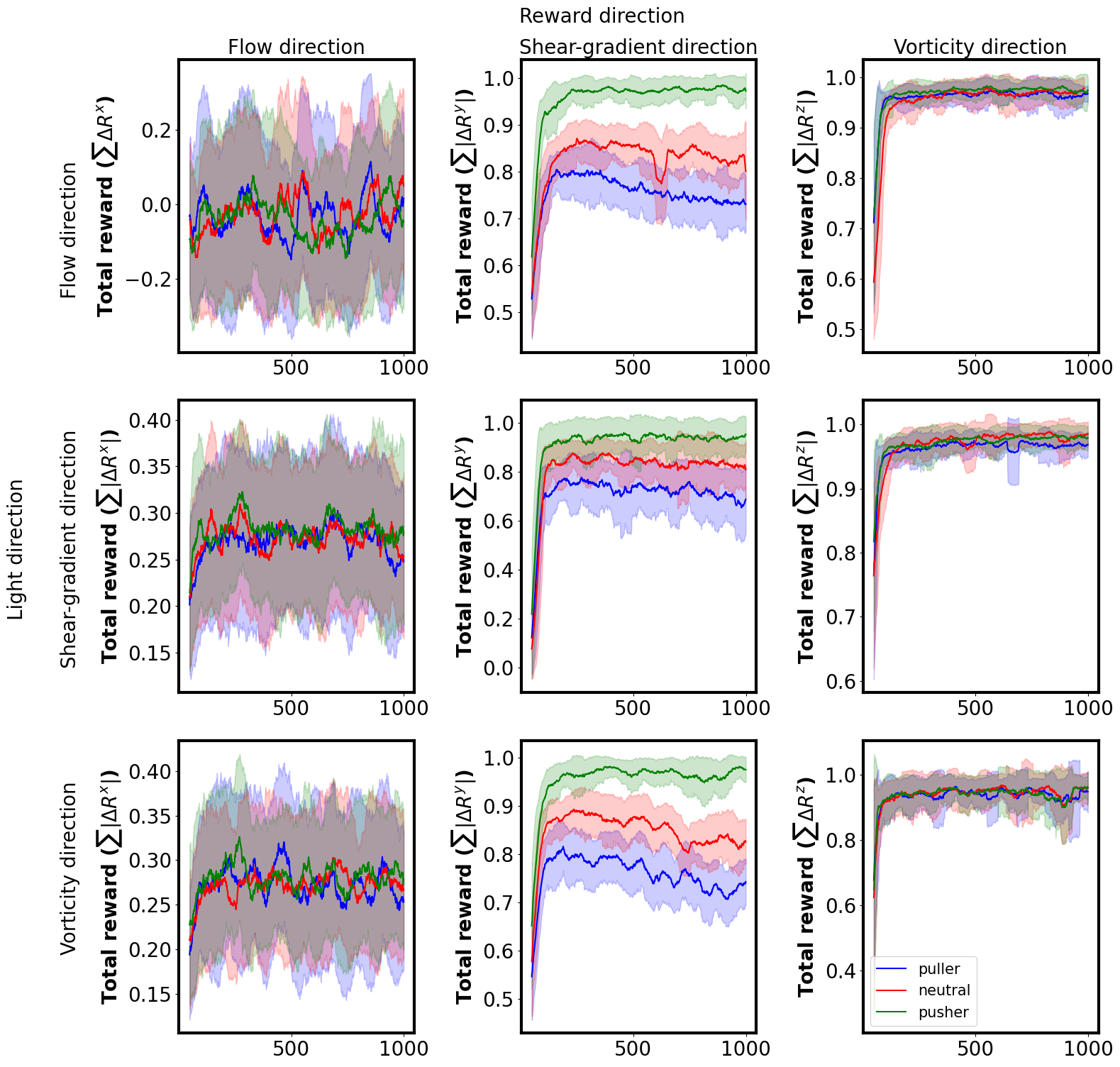}
    \caption{The rolling average of the normalized total rewards, with and averaging window of 50 episodes, for different types of swimmers, with eyes located at  $\Theta = 30^\circ$, using $\{\bm{\tau}_i,\hat{\bm{n}}_e\cdot\hat{\bm{n}}_L\}$ as the state defining variables. Here we consider three distinct swimmer types, puller ($\alpha = 2$), neutral($\alpha = 0$) and pusher($\alpha = -2$).}
    \label{fig5}
\end{figure*}

Fig.~\ref{fig3} shows the learning results for such a model swimmer. Each column of panels in this figure represents a reward direction, i.e. the desired swimming direction, and each row represents the direction of the signal source. In cases where the desired/target swimming direction is aligned with the signal direction, we use a signed reward ($\Delta R^\alpha(T_I)$), otherwise we use the corresponding unsigned reward ($\abs{R^{\alpha}(T_I)}$).
The rationale behind this is that the swimmer is able to distinguish whether it swims towards or away from a light source, hence the use of signed rewards in that case. However, when swimming in the other directions, it would not be easily able to break symmetry. The rewards in these cases are therefore unsigned. Inspired by the fact that the eye location of particular biological microorganisms, e.g. \textit{Chlamydomonas},
%MST name it/them here --> (A) Fixed
is at $30$ degrees away from the front \cite{Kreimer2009}, we consider a micro-organism-inspired swimmer with $\Theta = 30^\circ$. This swimmer can perceive three types of signals: surface stresses, light alignment and light rotation (flashing). The results using these signals are found to be broadly similar to the results for swimmers able to use lab frame information, see Fig.~\ref{fig1}. In Fig.~\ref{fig3}, two different combinations of input signals were studied, i.e. $\{\bm{\tau}_i,\hat{\bm{n}}_e\cdot\hat{\bm{n}}_L,(\hat{\bm{n}}_e \times \bm{\Omega})\cdot\hat{\bm{n}}_L\}$ and $\{\bm{\tau}_i,\hat{\bm{n}}_e\cdot\hat{\bm{n}}_L\}$ alone. One can see that there is little qualitative difference between these two sets of parameters. Thus, the surface stresses and the alignment with the light provides sufficient information to allow for efficient swimming. It is somewhat surprising that the swimmer learns to swim so well in the vorticity and shear-gradient directions. However, as for the swimmers with access to lab frame information, these swimmers are still unable to efficiently swim with the shear flow. 

To clarify what role, if any, is played by the positioning of the eye, we have compared the learning efficiency for different locations, $\Theta = 0^\circ, 30^\circ \text{and}\, 90^\circ$, with $\{\bm{\tau}_i,\hat{\bm{n}}_e\cdot\hat{\bm{n}}_L\}$ as the input signals. The results are shown in Fig.~\ref{fig4}, where a similar performance is obtained in all cases. Thus, for the tasks considered here, the choice of the eye's location seems to be relatively unimportant. 

Finally, we also compare the performance for different swimmer types by considering the squirming parameters $\alpha = -2$ (pusher) and $\alpha = 0$ (neutral), in addition to the puller studied in the rest of this article. 
%MST right? --> (A) Yes
The plots in Fig.~\ref{fig5} show the results obtained when using the surface stresses and light-alignment signals $\{\bm{\tau}_i,\hat{\bm{n}}_e\cdot\hat{\bm{n}}_L\}$. Here we found a surprising result for the task of swimming in the shear-gradient direction, in which there is a clear differences between the policy efficiency developed under different swimming modes. The pushers achieve the best performance, followed by neutral swimmers, with pullers being the least efficient. This distinction between swimming modes has also been observed in a confined system \cite{Oyamaa,lauga2013}. 
%The other two swimming tasks show no distinction among swimming modes, when learning to swim in the shear-gradient direction
\vspace*{-8pt}
\section{Conclusions}
We have performed direct numerical simulations, using the smooth profile method, coupled with a deep reinforcement learning algorithm to investigate the learning performance of a swimmer under an applied zigzag flow. We  considered three different swimming assignments, in which the swimmer is tasked with moving in the shear-flow ($x$), shear-gradient ($y$), or vorticity ($z$) directions. We demonstrated how different state information provided to the swimmer during the learning could result in vastly different performance. We studied the learning in cases where the swimmer receives either lab frame or body frame (local) variables. For the former, an efficient policy for migrating in the vorticity and shear-gradient directions emerged for swimmers given only their instantaneous orientation and the memory of the last two actions. However, for the task of swimming in the flow direction, the swimmer was unable to develop an efficient policy, meaning it could not target regions in which the flow was maximal. 
%MST I just want to pint out that this may  be very different for sinusoidally varying shear gradients!
For swimmers more closely inspired by microorganisms, e.g. copepods,
%MST name this earlier! --> (A) fixed
we assumed that the swimmer has six force/stress-sensing channels distributed on its surface, allowing it to sense relative differences with the local fluid velocity. This swimmer was also assumed to possess a sensor, e.g. a crude eye or photo-receptor that can sense light. Thus, the model micro-organism can also measure it's alignment relative to the direction of a light source, and perhaps also the flashing of this light due to it relative rotation. We found that providing  hydrodynamic forces, along with the light alignment signal, was sufficient to develop efficient strategies to perform the swimming tasks. In particular, we observed similar efficiency to the case of swimmers trained on (global) lab frame information. Additionally, we found that the location of the eye was inconsequential to achieve these swimming tasks, with this particular flow field.  We also investigated the differences in learning as a function of type of swimmer, by comparing pullers, pushers, and neutral swimmers. The pushers outperform the other two modes, with neutral swimmers better than pullers. We hope that our work may help motivate future studies on efficient swimming strategies for active particles and also further our understanding of model biological swimmers. One of the remaining challenges to address relates to our use of an external torque to define the action of the swimmer. Thus, while our swimmers are force-free, they are not torque-free. This might be reasonable for certain  artificial swimmers, but most biological micro-swimmers are both force-free and torque-free. In this future work we will consider learning under torque-free conditions.
%MST why not force free too? --> (A) The system is already force free 

\section*{Acknowledgements}
This work was supported by the Grants-in-Aid for Scientific Research (JSPS KAKENHI) under grant no. JP 20H00129, 20H05619, and 20K03786 and the SPIRITS 2020 of Kyoto University. 
R.Y. acknowledges helpful discussions with Profs. Hajime Tanaka and Akira Furukawa. MST acknowledges funding from Warwick University's Turing AI fund.

%%%END OF MAIN TEXT%%%

%\subsection{Animations}

% tables should appear as floats within the text
% The ruledtabular enviroment adds doubled rules to table and sets a
% reasonable default table settings.
% Use the table* environment to get a full-width table in two-column
% Add \usepackage{longtable} and the longtable (or longtable*}
% environment for nicely formatted long tables. Or use the the [H]
% placement option to break a long table (with less control than 
% \begin{table}%[H] add [H] placement to break table across pages
% \begin{ruledtabular}
% Lines of table here ending with \\
% \end{tabular}
% \end{ruledtabular}
% \end{table}

% Surround table environment with turnpage environment for landscape
% table
% \begin{table}
% \begin{ruledtabular}
% \end{tabular}
% \end{ruledtabular}
% \end{table}

%\appendix
%\section{}

%\begin{acknowledgments}
%\end{acknowledgments}
%\bibliographystyle{apsrev4-2}
\bibliography{manuscript.bib}

%\clearpage
%\begin{widetext}
%\input{supplement/supplememt.tex}
%\end{widetext}

\end{document}

\title{Supplemental Information of Learning to swim efficiently in a nonuniform flow field}
\maketitle
\section{SI1. System parameters}

In the main text, we introduced the systems parameters which characterize the behaviour of our system. Among these, the principle (adimensional) parameters are the particle radius $\sigma/\Delta=5$, the first squirming mode $B_1\rho_f\Delta/\eta = 0.1$, the height of the simulation box (along the shear-gradient direction) $L_y/\Delta = 64$, the shear rate $\dot{\gamma}\rho_f\Delta^2/\eta = 0.04$, the magnitude of the applied external torque $H\rho_f/(\eta^2\Delta) = 400$, the simulation time step $\Delta t\eta/(\rho_f\Delta^2) = 0.0714$, and the total number of simulation time steps (duration) per episode $T_{episode} = 2 \times 10^5$. With these, we can determine the particle Reynolds numbers (due to shear and swimming) as: 
\begin{align*} 
&\textbf{Reynolds number}\\
Re_{shear}&= \frac{\sigma^2 \dot{\gamma} \eta}{\rho_f} = 25(0.04) = 1.0 \\ 
Re_{swim} &= \frac{2}{3}B_1 = \frac{2}{3}(0.1) \approx 6 \times 10^{-2} \\ 
\end{align*}
The state of the system can be conveniently described in terms of the following non-dimensional parameters:
\begin{align*}
&\textbf{Systems parameters}\\
\psi_1 &= \frac{v_{swimmer}}{v_{shear}} = \frac{\frac{2}{3}B_1}{\dot{\gamma}\frac{L_y}{2}} \approx 5 \times 10^{-2} \\ 
\psi_2 &= \frac{\delta_{external}}{\delta_{shear}} = \frac{\frac{H \Delta t}{8\pi \sigma^3\eta}}{\frac{\dot{\gamma}}{2}\Delta t} \approx 6.3  \\ 
\psi_3 &= \frac{v_{swimmer}T_{episode}}{L_y}   \approx 21 \\ 
\psi_4 &= \delta_{external}T_{episode}    \approx 180 \\ 
\psi_5 &= \delta_{shear}T_{episode}       \approx 28 \\ 
\end{align*}
where the $\psi_2$ measures the ratio between the rotational velocity created by the external torque over the rotational velocity created by the shear on the swimmer, $\psi_3$ measures how many shear steps the particle can swim across over an episode, $T_{episode}$ ($\Delta t$ the simulation time step).

\section{SI2. Learning to swim in the flow direction using global information}
We have discussed how the choice of input signals affects the learning for the task of swimming in the flow direction using the global information. The plot included in the main text shows the reward over the training episodes. Here we present data in the form of violin plots. The white circles inside each ``violin'' correspond to the average normalized total reward, evaluated as the average displacement in the flow direction. The thick vertical line represents the interquartile range. In the last violin plot, one can see that giving information of the global location ($\bm{R}$ and the instantaneous orientation ($q$), seems to yield better performance, with the distribution shifting towards higher values. However, more extensive simulations are required to validate this. In any case, the swimmer still experiences difficulty in aligning itself with the flow direction (the overall performance is still quite poor).
\begin{figure}[H]
    \centering
    \includegraphics[trim={0.3cm 0 0 0.2cm},clip,width=0.8\textwidth]{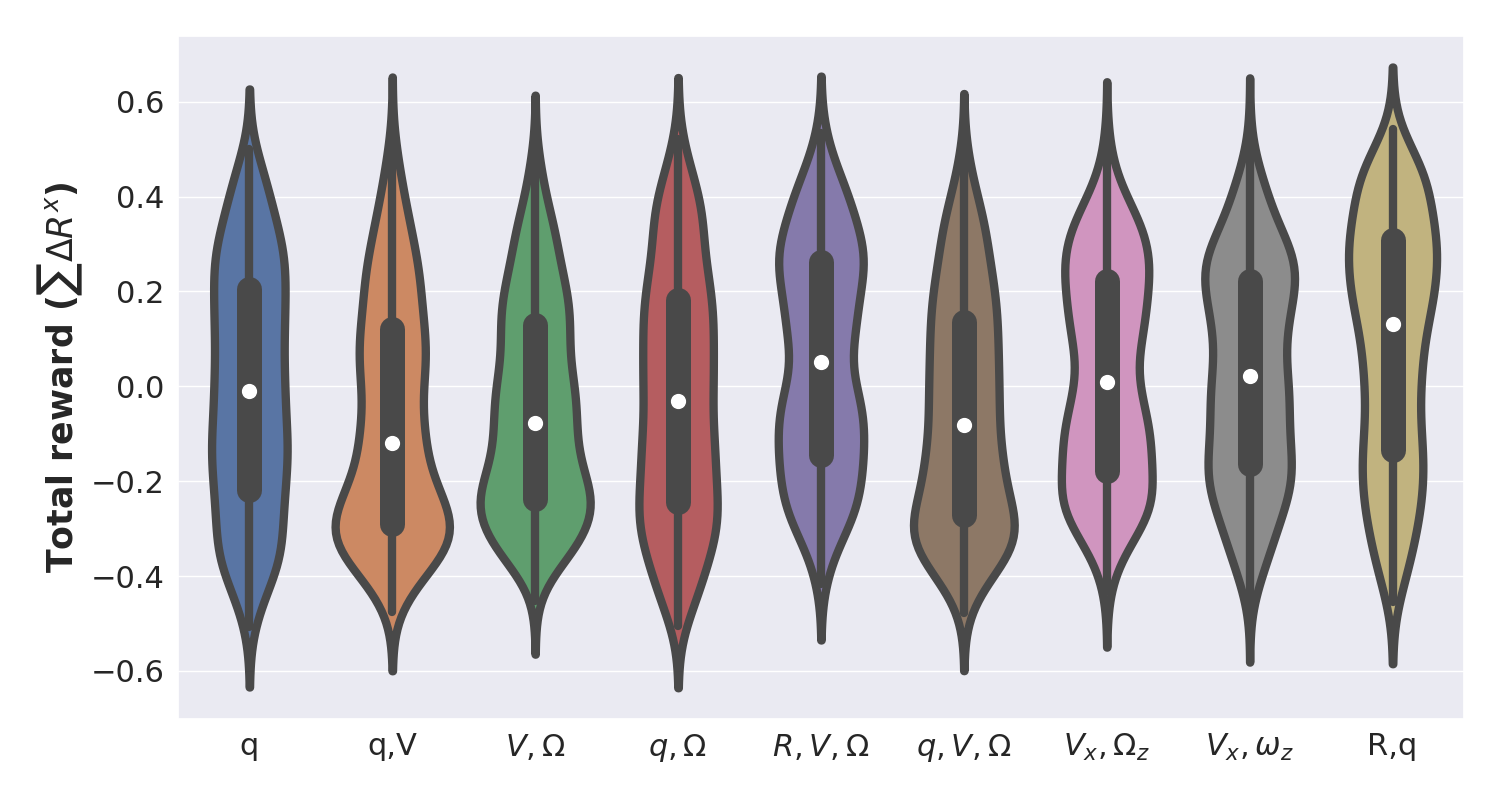}
    \caption{Violin plots for various sets of input variables, i.e. quaternion ($q$), translational velocity ($\bm{V}$), rotational velocity ($\bm{\Omega}$), swimmer location in the lab-frame of reference ($\bm{R}$) and the background shear vorticity ($\bm{\omega}$). All sets include the last two taken actions ($a$).}
    \label{figSI5}
\end{figure}